\documentclass[fleqn,usenatbib]{mnras}

\usepackage[utf8]{inputenc} 

\usepackage{newtxtext,newtxmath}

\usepackage[T1]{fontenc}

\DeclareRobustCommand{\VAN}[3]{#2}
\let\VANthebibliography\thebibliography
\def\thebibliography{\DeclareRobustCommand{\VAN}[3]{##3}\VANthebibliography}

\usepackage{graphicx}	
\usepackage{amsmath}	
\usepackage{amssymb}	

\title[Self-Lensing XRBs]{Self-Lensing in Eclipsing X-ray Binary Systems}

\author[Sorabella et al.]{
Nicholas M. Sorabella,$^{1}$\thanks{E-mail: Nicholas$\_$Sorabella@student.uml.edu}
Silas G.T. Laycock,$^{1}$
\\
$^{1}$Dept. of Physics and Applied Physics, UMass Lowell, 660 Suffolk St, Lowell, MA 01854\\
}

\date{Accepted XXX. Received YYY; in original form ZZZ}

\pubyear{2020}

\begin{document}
\label{firstpage}
\pagerange{\pageref{firstpage}--\pageref{lastpage}}
\maketitle

\begin{abstract}
Abstract: This project examines the feasibility of using gravitational lensing to measure the mass of compact objects in eclipsing X-ray binary (XRB) systems. We investigate which kind of XRB would be most conducive for viewing the effect, by modeling the amplification curves and determining if any feature of an XRB system could potentially hinder observation of such a signal. We examine the effect of accretion disks and stellar winds, as well as the compact object mass, binary separation, and companion spectral type. Generally speaking, the lensing signal is strongest for when the angular size subtended by the companion is small, favoring relatively compact companion stars (LMXBs) although evolved massive stars (such as certain WR stars) have signals that are feasibly detectable. Interestingly, the self-lensing signal is stronger in binaries with large separations, which is the exact opposite of the case for all other techniques. Thus, a dedicated self-lensing survey would complement X-ray and radial-velocity techniques, by extending the parameter space for discovery of compact objects. Simultaneously, a self-lensing survey offers the possibility of revealing the large population of non-accreting compact objects in galactic binary systems.
\end{abstract}




\section{Introduction}

The high mass X-ray binary (XRB) system IC 10 X-1 has been studied for many years by several groups in the X-ray astronomy field. Initial findings \citep{Prestwich_2007,Silverman_2008} suggested that the massive Wolf-Rayet (WR) star in the system had an equally massive compact companion: a black hole of about twenty-four solar masses. The system appeared to be an eclipsing binary where the X-rays from the black hole would periodically dip as the WR star passed in front of it from our perspective on Earth. The period of these dips, coupled with the measured optical radial velocity of the WR star, produced the mass function that included the possibility that this was the most massive black hole discovered in an XRB. However, further research showed that the phase of the radial velocity curve did not match up with the observed eclipses in the X-ray light curve \citep{2015MNRAS.446.1399L}. It was instead suggested that the radial velocity that was measured was not of the star itself but a feature in the dense wind of the star. It was this fact that put the mass of the black hole into question. The compact object could still be the behemoth it once was speculated to be, or it could be a less massive neutron star instead. With the mass function of the compact object now unknown, the question that arose was “what other way can the mass of a compact object in an eclipsing binary system be measured”?

Black holes and neutrons stars are the most compact objects known to science. They are considered to be the best laboratories for researching General Relativity due to their extreme gravity. The space-time around these objects is warped so greatly that they can act as a lens in space \citep{Einstein506}. 
\begin{figure} 
    \centering
    \includegraphics[width=\columnwidth]{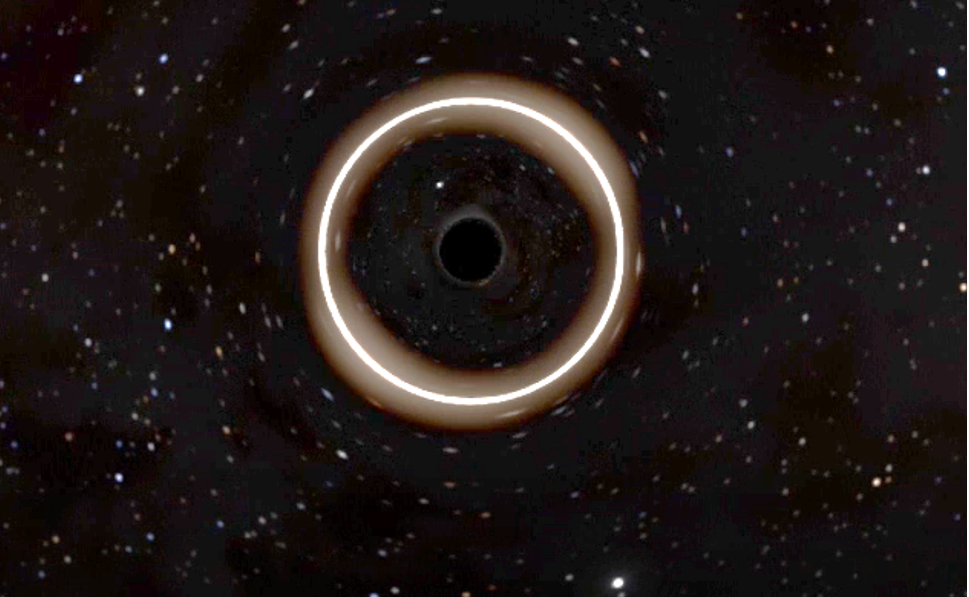}
    \caption{A black hole gravitationally lenses a background star. This simulation was created using the software SpaceEngine \citep{romanyuk2016space}.}
    \label{fig:1}
\end{figure}
\begin{figure*}
    \includegraphics[width=6.95in]{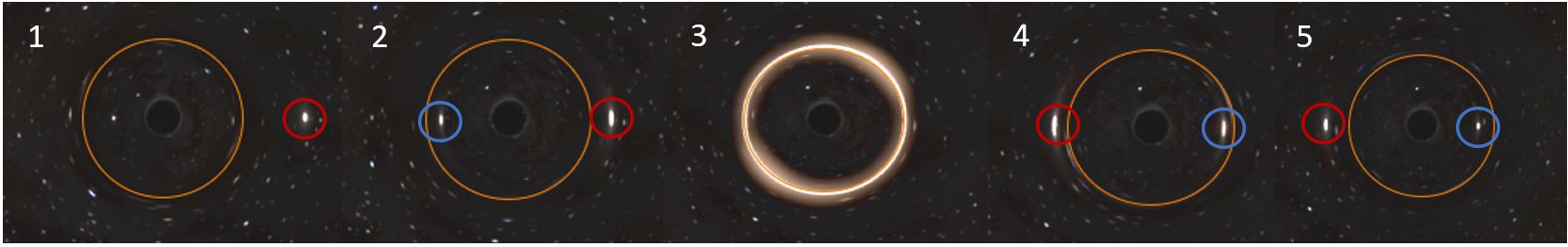}
    \label{fig:2}
    \centering
     \caption{Here we see a step by step process of a black hole gravitationally lensing a background star. Panel 1 shows the star (circled in red) just outside the Einstein Radius (orange). Panel 2 shows the surface area of the star increasing as a secondary image of the star (circled in blue) forms. Panel 3 depicts when the black hole is directly in front of the star. Here the two images of the stars connect to form a perfect ring around the black hole where max amplification occurs. Panel 4 shows the image of the star returning to normal as the secondary image begins to fade. Panel 5 shows the lensing process completed (created using SpaceEngine \citep{romanyuk2016space}).}
\end{figure*}
The image of a distant object behind a black hole is both distorted and magnified, making it appear bigger and brighter than it would normally appear. We see this effect in Figure 1 and Figure 2 which were created using the simulation software SpaceEngine developed by \citet{romanyuk2016space}. This effect is known as gravitational lensing and we have seen many examples of such events in stellar light curves
(often called microlensing) \citep[e.g.][]{1993Natur.365..621A}.

This paper explores the idea of measuring the mass of compact objects in eclipsing XRB systems by exploiting this phenomenon. In eclipsing binary systems where the components are two stars, we see dips in the stellar light curve that tell us when one star has passed in front of the other. \citet{1969ApJ...156.1013T} were the first to mention that in an eclipsing binary system where at least one of the components is a compact object, such as a neutron star or black hole, there will not be a dip as expected. Instead, due to gravitational lensing, there will be a small increase in the apparent magnitude of the star. From this pulse, which would periodically happen every time the compact object passes in front of the star, as suggested by \citet{1973A&A....26..215M}, we would be able to deduce the mass of the compact object and the inclination angle of its orbit.
 
This effect (known as self-lensing) has only been measured a handful of times. \citet{Kruse275} were the first to observe the effect using the Kepler Space Telescope. The Quasiperiodic Automated Transit search algorithm discovered a series of periodic increases in brightness in the star as opposed to the dips expected by a planet \citep{Carter_2013}. The amplification was only a factor of 1.001, a relatively small increase but still a significant detection in the Kepler data (see \textit{Figure 2} of their paper). From this increase in brightness, together with the duration and profile of the feature, the researchers inferred that the object was not a planet but, rather, a white dwarf (WD) star of around 0.63 solar masses \citep{Kruse275}. Further self-lensing discoveries have all been found in the same way; all five self-lensing systems, so far, were found through Kepler data and consisted of a WD star eclipsing its sun-like companion \citep{2020arXiv200104448M}.

\section{Self-Lensing Theory}
The mathematics of this self-lensing effect have been developed over the years. \citet{1994ApJ...430..505W} derived a mathematical description of gravitational lensing for extended sources,
\begin{figure} 
    \centering
    \includegraphics[width=\columnwidth]{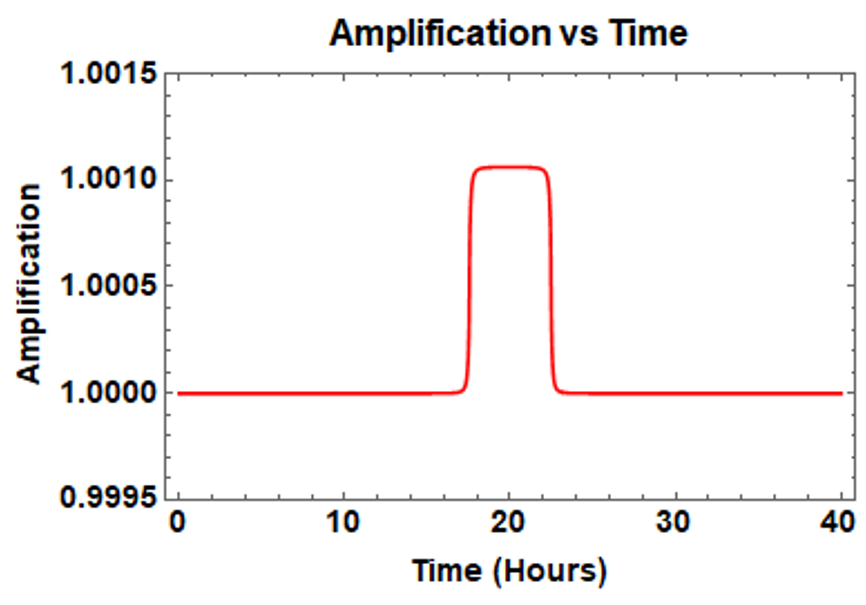}
    \label{fig:3}
    \caption{Simulated amplification curve for KOI-3278. The simulated curve is less rounded than the observed pulse, \citep{Kruse275}, as we do not account for limb-darkening. However, the maximum amplification and timescale of the event do not deviate in any significant manner and will be adequate for the focus of this paper.}
\end{figure}
as opposed to the point sources most microlensing events are treated as, with constant surface brightness. \citet{2011MNRAS.410..912R} developed the mathematics for binary systems specifically, though his work focused exclusively on the lensing of main sequence stars by their compact companions. Using the systems of equations published by both groups, we have developed a code that can simulate lensing events in binary systems given the mass of the star and compact object. Figure 3 shows our simulated amplification curve for KOI-3278. It is worth noting that the shape of the observed signal is slightly more rounded than our theoretical model. This rounding is due to limb-darkening where, as \citet{Han_2016} described, a limb-darkened stellar disk will produce a more rounded self-lensing pulse. \citet{Han_2016} also showed that the max amplification seen on a limb-darkened disk where the lens is located at the center of the source is slightly higher than the magnification seen for a disk with constant surface brightness. Simultaneously, \citet{Han_2016} showed that a limb-darkened disk will produce lower magnification than a uniform disk if the lens is more inclined from the center of the source star. According to \citet{1994ApJ...430..505W}, limb-darkening can account for a 10-15$\%$ discrepancy in the maximum amplification observed. Given that this change is within the same order of magnitude as assuming constant surface brightness, we chose the latter as to not assume a limb-darkening coefficient for all XRBs. We wrote our code in the program Mathematica where, given the mass of the star and compact object, it will solve for the amplification curve as a function of time. The code first evaluates for the orbital distance of the compact object using the equation
\begin{equation}
 a=\sqrt[3]{\frac{P^2G(M+m)}{4\pi^2}}
\end{equation}
where $a$ is the semi-major axis, $P$ is the period of the orbit (known through the eclipses) and $M$ and $m$ are the masses of the compact object and star respectively. From there we solve for the angular velocity $\omega$ which can be used to find the transverse velocity $v_{\perp}$.
\begin{equation}
  \omega=\sqrt{\frac{G(M+m)}{a^3}}  
\end{equation}
\begin{equation}
  v_{\perp}=a\omega
\end{equation}
Next, we have to solve for the Einstein Radius ($R_E$) using the Schwarzschild Radius ($R_s$) of the compact object.
\begin{equation}
    R_{s}={\frac{2GM}{c^2}}
\end{equation}
\begin{equation}
    R_{E}={\sqrt{2R_{s}a}}
\end{equation}
From here we can solve for the impact parameter $P_{\star}$, which is the ratio between $R_E$ and the radius of the star $R_{star}$.
\begin{equation}
    P_{\star}= {\frac{R_{star}}{R_{E}}}
\end{equation}
The Einstein Time $t_{E}$, the time it takes the compact object to move the length $R_{E}$, is found using
\begin{equation}
    t_{E}={\frac{R_{E}}{v_{\perp}}}
\end{equation}
which leads to our equation for angular separation in units of the angular Einstein Radius
\begin{equation}
    u(t)={\sqrt{u_{0}^2+\left({{\frac{t-t_{0}}{{t_{E}}}}}\right)^2}}
\end{equation}
given
\begin{equation}
    u_{0}=\frac{a}{R_{E}}\phi
\end{equation}
where $u_{0}$ is the closest angular approach and $\phi$ is the inclination of the orbit. Lastly, the variable $P_{\star}$ and the function $u(t)$ can be plugged into our equation for amplification versus time given by
\begin{equation*}
    \begin{aligned}
    A[u,P_{\star}]=&\frac{1}{2\pi}\left[{\frac{u+P_{\star}}{{P_{\star}^2}}\sqrt{4+(u-P_{\star})^2}}E(k)\right.\\
    &-\frac{u-P_{\star}}{P_{\star}^2}\frac{8+u^2-P_{\star}^2}{\sqrt{4+(u-P_{\star})^2}}K(k)\\
    &+\left.{\frac{4(u-P_{\star})^2}{P_{\star}^2(u+P_{\star})}\frac{1+P_{\star}^2}{\sqrt{4+(u-P_{\star})^2}}\Pi(n;k)}\right]
    \end{aligned}
\end{equation*}
where $E(k)$, $K(k)$ and $\Pi(n;k)$ are the complete elliptic integrals of the first, second and third kind, respectively, as defined by \citet{gradshteyn2014table} and $n$ and $k$ are given by
\begin{equation}
    n=\frac{4uP_{\star}}{(u+P_{\star})^2}
\end{equation}
\begin{equation}
    k=\sqrt{\frac{4n}{4+(u-P_{\star})^2}}
\end{equation}
\citep{2011MNRAS.410..912R}.

\section{Self-Lensing in XRBs}
As it stands, no black hole or neutron star’s mass has ever been measured using self-lensing. Given that most (but not all) black holes and neutron stars discovered are in XRB systems, it is important to look at how feasible it would be to measure the effect in these types of systems. As shown above, the magnification from lensing is a function of the ratio between the radius of the star $R_{star}$, and the Einstein Radius $R_{E}$. XRB systems are, by definition, closely orbiting systems which would reduce the amount of magnification we would be able to see due to $R_{E}$’s dependence on $a$. It appears that the best systems to look for self-lensing are ones where the star is not a giant star. Most high mass X-ray binaries (HXMBs) like Cygnus X-1 are
home to extremely large and massive blue stars which, due to their size, would show a smaller relative flux increase than the WD example of KOI-3278. Interferingly however, as we see in Figure 4, although our predicted
\begin{figure} 
    \centering
    \includegraphics[width=\columnwidth]{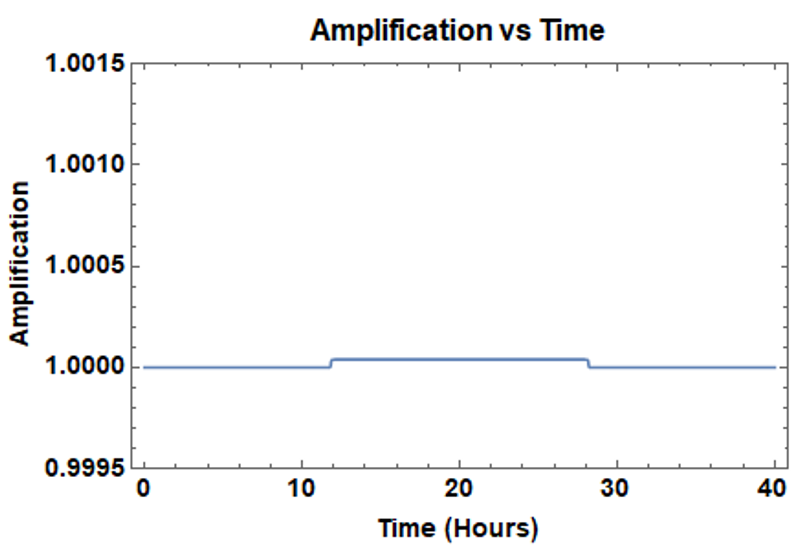}
    \label{fig:4}
    \caption{Simulated amplification curve of the XRB Cygnus X-1 if the system was treated as eclipsing. Notice the reduced amplification but longer time scale of the event. HMXBs containing main-sequence stars will produce minuscule relative flux increases due to the stellar companion's size relative to $R_{E}$. This model assumes 14.81$M_\odot$ for the black hole, 19.16$M_\odot$ and 16.42$R_\odot$ for its stellar companion, and an orbital period of 5.59 days as found by \citet{Orosz_2011}}
\end{figure}
signal for Cygnus X-1 (known to be a non-eclipsing system \citep{1985AZh....62..731L} but treated as such for a model example) shows a relative flux increase of 1.00004, its duration is much longer. Low mass X-ray binaries (LMXBs) would be better candidates due to the much smaller radius of the host star allowing for much higher potential magnification. However, the likelihood of eclipses are far smaller in LMXB's given that the donor subtends a far smaller angle to the compact object as a shorter period HMXB \citep{charles1995exploring}. Simultaneously, LMXBs often have large accretion disks due to the fact that they are actively feeding on material that comes directly from their host stars. Depending on the size, it is entirely possible that at certain wavelengths the disk blocks or distorts the lensing signal by partially obscuring the star. However, HMXBs feed off the stellar wind of their host stars and not the star itself which produces a much smaller accretion disk than in roche-lobe fed systems. Simultaneously, the Einstein Radius (where lensing occurs) is a function of the mass of the compact object and its distance from its host star. In the case of IC 10 X-1, the smallest possible Einstein Radius would be 13,310 km, far larger than the accretion disk would be. This would indicate that, in the case of HMXBs, their accretion disks would likely not block any lensing and, by extension, not affect the amplification curves. 

The best XRB system to look for a self-lensing event would be one that combines the best characteristics of both HMXB and LMXB systems. This would be best personified in an HMXB system containing a WR star such as IC 10 X-1 or Cygnus X-3. These systems are thought to harbor the most massive stellar black holes \citep{Silverman_2008}, while the Wolf-Rayet star having lost its hydrogen envelope, is much smaller than any other star of comparable mass. According to stellar evolutionary models, a WR star with a mass of 30$M_\odot$ can have a radius of only about 2$R_\odot$ \citep{1989A&A...210...93L}. A compact companion would be feeding off the stellar wind of the WR star just as it would with any other HMXB system meaning it would have an accretion disk that is much smaller than its LMXB counterpart. These systems are also likely to have shorter orbital periods on the order of a few hours to a few days. This would make the acquisition of repeated signals much more likely and allow for signal averaging over time.

The WR star in the IC 10 X-1 system has a most probable mass of 34$M_\odot$ meaning, according to the tabulated WR mass-radius relation from \citet{1989A&A...210...93L}, it possesses a radius of $<$2$R_\odot$. The period of the system is known to be 34.9 hours, but, as mentioned previously, the mass of the compact object is essentially unknown \citep{2015MNRAS.452L..31L}. Figure 5 shows our simulated amplification curves for different mass black holes for the IC 10 X-1 system. As one can see, the max amplification is of similar magnitude to those seen in the WD examples. This would imply that it should be possible, given the current technology available, to view such an event in a system like IC 10 X-1.

The inclination of the compact object’s orbit will have noticeable effects on the overall shape of the amplification curve.
\begin{figure} 
    \centering
    \includegraphics[width=\columnwidth]{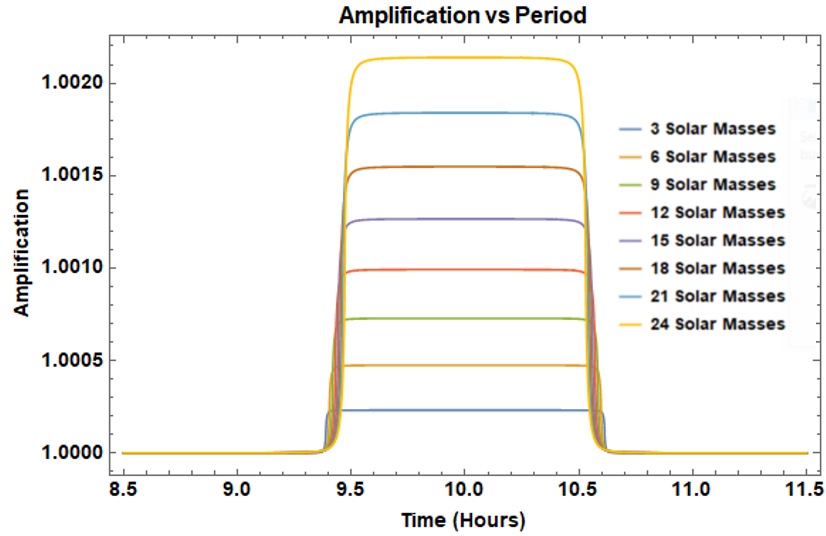}
    \label{fig:5}
    \caption{Simulated amplification curves for numerous mass black holes for IC 10 X-1. We assume 34$M_\odot$ and 2$R_\odot$ for the WR star and an orbital period of 34.9 hours. The maximum amplifications seen are of similar magnitude as that of all WD self-lensing events.}
\end{figure}
\begin{figure} 
    \centering
    \includegraphics[width=\columnwidth]{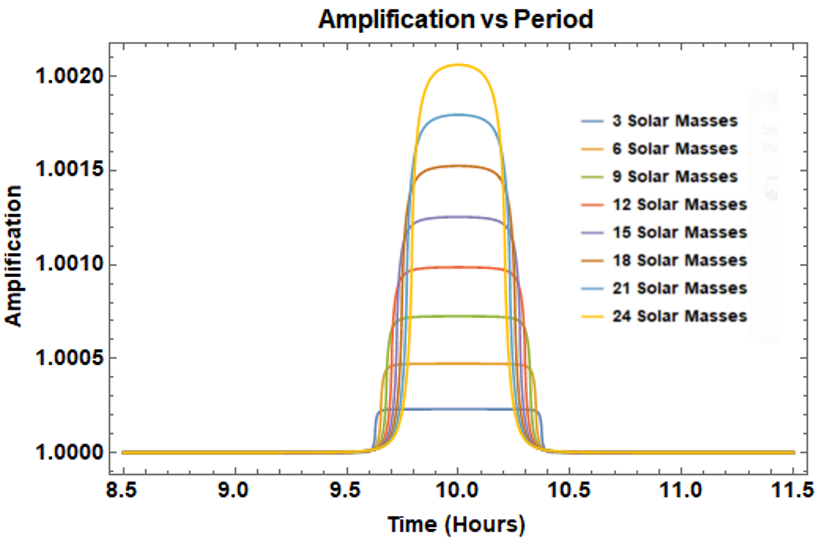}
    \label{fig:6}
    \caption{Simulated amplification curves for numerous mass black holes for IC 10 X-1 with an orbit that is inclined by 5 degrees from the center of the star. The maximum amplification remains the same but the timescale of the event is reduced.}
\end{figure}
However, in the case of XRB systems, we see that a larger inclination angle does not affect the max amplification seen on the curve. Instead, only the length of the event is reduced. Figure 6 demonstrates this feature. This means, with the combination of the max amplification and the overall length of the lensing event, one could deduce both the mass of the compact object and the inclination of its orbit. This is an important
\begin{figure} 
    \centering
    \includegraphics[width=\columnwidth]{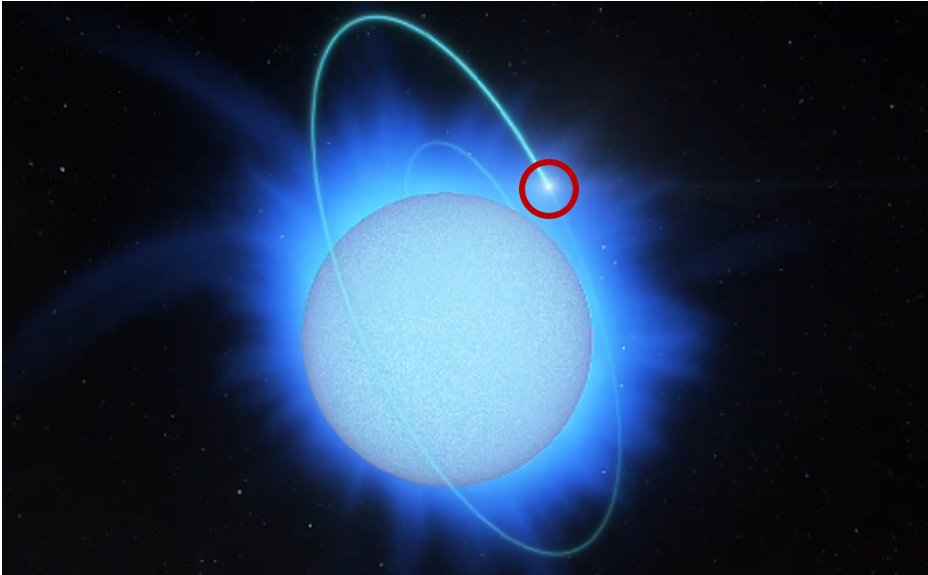}
    \label{fig:7}
    \caption{A simulated example of a black hole (circled in red) being eclipsed by the wind of its host star (created using SpaceEngine \citep{romanyuk2016space}). This scenario is likely for the IC 10 X-1 system where the optically thick wind blocks the X-rays originating near its compact companion.}
\end{figure}
feature for IC 10 X-1 as, in the case of that specific system, it is not certain that the compact object eclipses the star as once thought \citep{2015MNRAS.452L..31L}. Instead, it is entirely possible that only the wind of the star blocks the X-rays emitted from the compact object as seen in Figure 7 (created using SpaceEngine). However, self-lensing can be used to determine whether this is the case or not. 
\begin{figure} 
    \centering
    \includegraphics[width=\columnwidth]{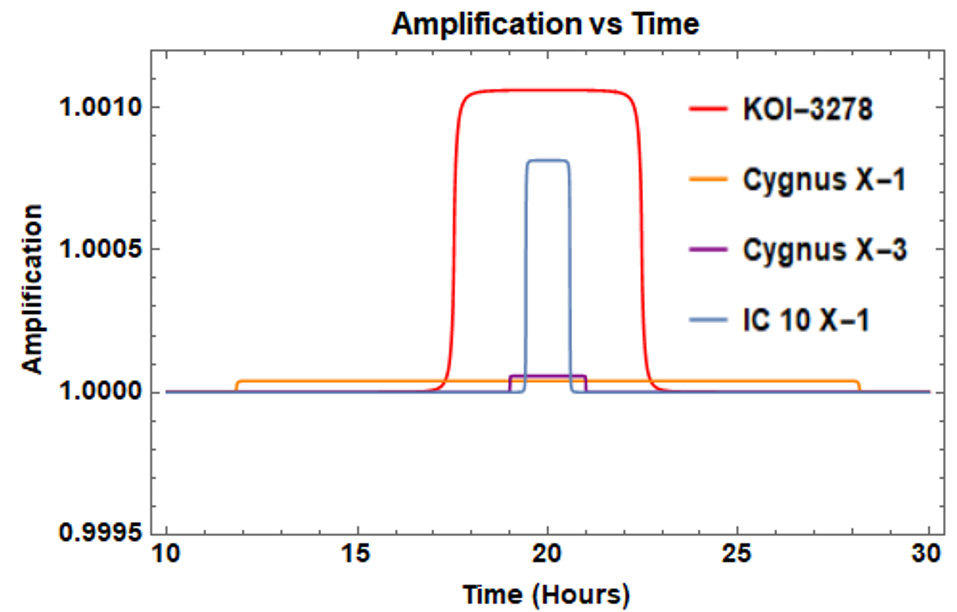}
    \label{fig:8}
    \caption{Simulated amplification curves for 3 different HMXB systems compared with the simulated KOI-3278 signal. Notice that, due to its far tighter orbital radius, Cygnus X-3 produces a far smaller max amplification than IC 10 X-1 even though both host stars are of similar size.}
\end{figure}
The simulated amplification curves from figures 5 and 6 show that the amplification expected from this system are of similar magnitude to that of the white dwarf system. If no amplification were seen in the system then this would most likely mean that lensing does not reach the surface of the star and, thus, the compact object must not eclipse the star. For IC 10 X-1, this would be around 7 degrees inclined from the center of the star.

Figure 8 models the theoretical amplification signal of several real systems representing different types of HMXBs in comparison to KOI-3278. We have already discussed IC 10 X-1 at length (in this model we assume a 10$M_\odot$ black hole) but we have yet to discuss the parameters of the others. Cygnus X-1 is a non-eclipsing prototypical HMXB home to what is likely a 14.81$M_\odot$ black hole and 19.16$M_\odot$ O-type star with an orbital period of 5.59 days \citep{Orosz_2011}. Even with a reasonably long orbital period by HMXB standards, the sheer size of its O-type companion (16.42$R_\odot$) results in a rather small amplification of only 1.00004. Like IC 10 X-1, Cygnus X-3 contains both a compact object (either a higher mass neutron star or lower mass black hole) and a WR star. However, even with the reduced radius of the WR star, the max amplification is not only much smaller than KOI-3278 but is barely larger than Cygnus X-1! The reason for this lies in the period of Cygnus X-3. Unlike IC 10 X-1 which orbits every 34.9 hours, Cygnus X-3 has an orbital period of just 4.8 hours \citep{2017MNRAS.472.2181K}. The impact parameter, used to find the max amplification, is the ratio between the radius of the star and the Einstein Radius. Given that the Einstein Radius is also a function of the semi-major axis, a shorter orbital period will result in a smaller $R_{E}$ and, thus, smaller amplification. With that in mind, it is important to look at both the mass of the compact object as well as its period.

\section{Orbital Period's Effect on Amplification}
We can plot the max amplification for black holes in HMXBs containing WR stars for a variety of different mass black holes and orbital periods.
\begin{figure} 
    \centering
    \includegraphics[width=\columnwidth]{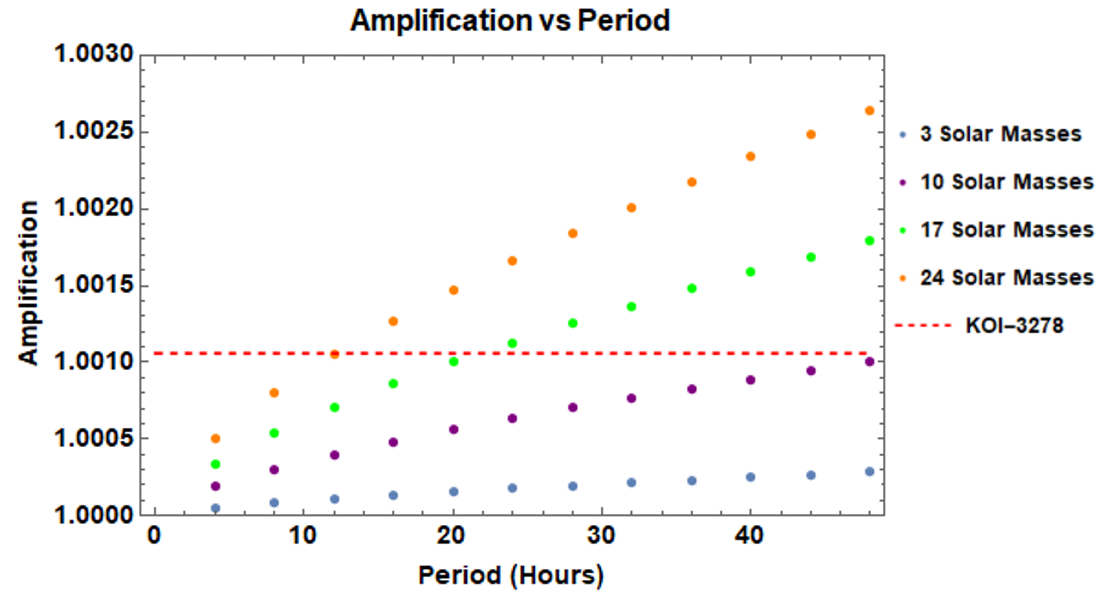}
    \label{fig:9}
    \caption{Maximum amplification for various mass black holes in HMXBs containing WR stars versus period as compared with KOI-3278. Notice that lower mass black holes need far larger orbital periods to produce the same relative flux increases as their higher mass counterparts.}
\end{figure}
Figure 9 shows the max amplification for 3, 10, 17, and 24 solar mass black holes at different orbital periods around the same WR star (assuming a stellar radius of 2$R_\odot$). For a baseline, the dashed red line represents the max amplification seen for KOI-3278, a value of similar magnitude to all observed WD self-lensing events. It is easy to see that, even with an orbital period of only 4 hours, the higher mass black holes are of similar magnitude to what was observed with KOI-3278. However, smaller black holes (also neutron stars) produce max amplification up to 20 times smaller at periods $<$5 hours. This is why Cygnus X-3 has such a smaller theoretical max amplification than IC 10 X-1. For a compact object such as Cygnus X-3, it would need a period about 6 times longer than that of a 10$M_\odot$ black hole to produce the same level of amplification. This means that a low mass black hole would need an orbital period of $>$12 days to have the same max amplification as KOI-3278. On the opposite end of the spectrum, a 24$M_\odot$ black hole with a WR companion can produce max amplifications many times larger than the WD systems if their orbital period is $>$48 hours.

\section{Circumventing the Wind of the Star}
All of our assumptions regarding WR systems are derived from the fact that they can have a reduced radius of $<$2$R_\odot$. In the case of IC 10 X-1, \citet{2015MNRAS.452L..31L} used the tabulated WR mass-radius relation from \citet{1989A&A...210...93L} to estimate this sub 2$R_\odot$ radius. However, \citet{2015MNRAS.452L..31L} also estimated a total radius of the star plus its intense wind to be around 8$R_\odot$ given the length of the observed X-ray eclipses. The wind is far larger than the actual surface of the star and the Helium inside the wind is ionized, leading to the emission of photons. In the visible spectrum, the wind far outshines the surface of the star, meaning, when calculating lensing at these wavelengths, $R_{star}$ should really be the radius of the wind or 8$R_\odot$.
\begin{figure}
\includegraphics[width=\columnwidth]{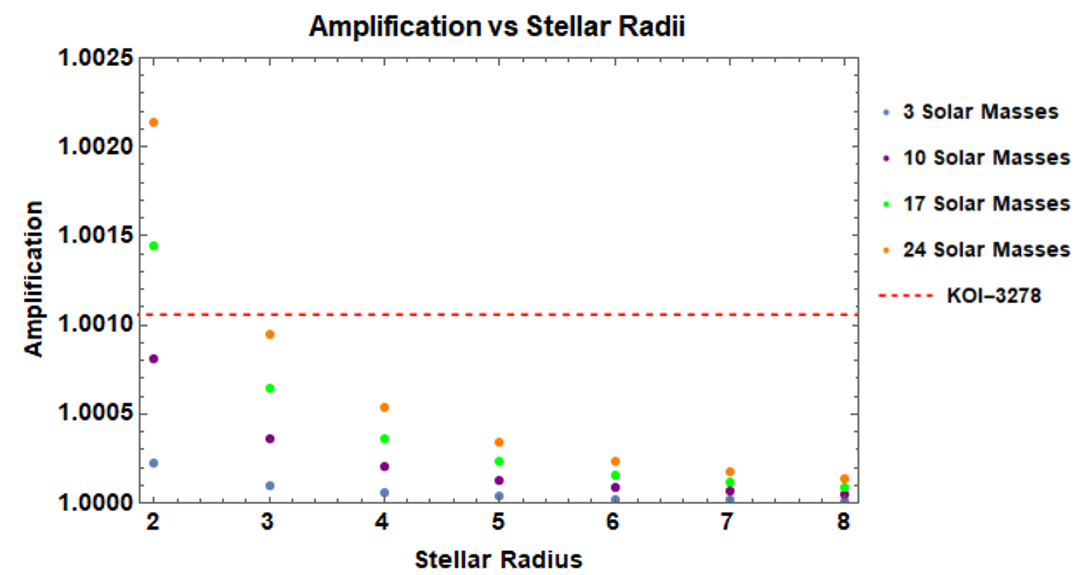}
\centering
\label{fig:10}
\caption{Maximum amplification for various mass black holes in HMXBs containing WR stars versus the radius of the star as compared with KOI-3278. Notice that even larger mass black holes produces max amplifications an order of magnitude lower than KOI-3278 at $>$8$R_\odot$.}
\end{figure}

Figure 10 details how changing the stellar radius affects the max amplification for various mass black holes given an orbital period of 34.9 hours. There is a large dip in max amplification from 2$R_\odot$ to 3$R_\odot$ where, by 3$R_\odot$, we see all possible max amplifications given are now less than that of KOI-3278. By 8$R_\odot$ (the estimated radius of the wind in IC 10 X-1) we see that even a 24$M_\odot$ black hole would produce a max amplification that is an order of magnitude less than KOI-3278. This would imply that a light curve in the visible spectrum of the system would show little amplification.

Knowing that the wind of the star could negatively affect the max amplification observed, it is important to explore options that could bypass the wind entirely. The most intuitive would be to take multiple light curves of the system using different filters and averaging out the effect from the wind. Another prospective method would be to observe the system in a longer wavelength that can easily cut through the wind and observe the star directly. \citet{1983ApJ...271..221H} discussed the possibility of examining WR stars in the infrared to peer through optically thick winds. As they suggested, for a star with an extended atmosphere, the radius at which thermal opacity has unit optical depth is highly wavelength dependent. Their model showed that, at 10$\mu$m, the effective radius was 15$R_\odot$ while at 2$\mu$m the radius was only 4.8$R_\odot$ (assuming $T= 3.3 * 10^4$ K) for the WN star they were studying \citep{1983ApJ...271..221H}. This would seem to indicate that there is a fine line in choosing the correct wavelength as, given too long of a wavelength, the observation would be focused on the outermost layers of the wind. However, as \citet{1983ApJ...271..221H} suggested, the optimal wavelength for viewing the interior of the wind would likely be around 2$\mu$m. Regardless, the choice of wavelength for observation will have a great effect in determining what relative flux would be observed and warrants further study.

\section{Future Observations}
While simulation of amplification curves is important, it is equally important to examine what telescopes can be used to measure a self-lensing event. LAMOST, an optical survey that has recently begun observation, can be used to mine for stellar-mass black holes using its sensitive radial velocity measurements \citep{2019ApJ...886...97Y}. What remains an interesting prospect for LAMOST is its potential to discover inactive XRB systems in our galaxy solely off of radial velocity measurements which would greatly expand the list of possible self-lensing candidates. 

All previous self-lensing events discovered have been made through the Kepler Space Telescope. This makes sense given that Kepler's true purpose was being sensitive enough to view eclipses of small planets around stars. In search of self-lensing events around XRB systems, TESS, Kepler's successor, would be an excellent candidate given that it is an all sky survey mission. A potential tactic would be using radial velocity data from LAMOST to find binary systems with black hole or neutron star candidates and TESS to check the system for self-lensing pulses \citep{2019ApJ...883..169M}. Assuming that the optimum band-pass for reducing the effective $R_{star}$ is in the IR, \citep{1983ApJ...271..221H}, JWST would make an excellent candidate for future infrared observations.     

\section{Conclusions}
Initial results show that self-lensing in XRB systems could be viewed given the correct conditions. The orbital parameters and relative size of the star in LMXBs could provide higher amplifications than HMXB systems but the size of the accretion disk could hinder any observation of a lensing event. HMXB systems possess compact objects with far smaller accretion disks but the sheer size of the massive star, combined with the tight orbital radius, leads to a far smaller max amplification. However, HMXBs containing WR stars, due to their reduced radius, show great promise for producing a measurable lensing event given a compact object of sufficient mass and orbital period. If a lensing event were measured in such a system, the max amplification and length of the event would reveal both the mass of the compact object and the inclination of its orbit. Self-lensing can also be used to further constrain the inclination of a binary system by revealing if the compact object eclipses the surface of the star or if the X-rays are eclipsed by the wind of the star instead. 

In systems containing powerful stellar winds, it is important to observe the star using different filters or longer wavelengths to reduce the effective $R_{star}$ and maximize potential amplification. \citet{1983ApJ...271..221H} showed that the optimal wavelength for peering through the WR wind would be around 2$\mu$m. This would indicate observations in the infrared (potentially by JWST) could provide the best chance of viewing a lensing event in such a system.  

The LAMOST survey has great potential in uncovering new stellar mass black holes in our galaxy and should be able to uncover inactive HMXBs and LMXBs. From LAMOST's findings, TESS can be used to examine these new systems for self-lensing events as suggested by \citet{2019ApJ...883..169M}. Overall, self-lensing may provide a unique method of mass measurement for XRB systems independent of any other technique currently used in astronomy.

\section*{Acknowledgements}

We would like to thank Sayantan Bhattacharya for reading the manuscript and offering advice regarding the wind of IC 10 X-1. Nicholas Sorabella would also like to thank David Simon for being the first person who listened to him explain this concept and encouraged him to pursue it.

\section*{Data Availability}
The only data that were used throughout this paper were the masses, radii, and orbital periods of various astronomical objects. These values were taken from the findings of various studies that were cited throughout the paper. The data used to find these values can be found in their respective papers. 




\bibliographystyle{mnras}
\bibliography{self_lensing_xrbs} 







\bsp	
\label{lastpage}
\end{document}